# Recovery Time Metric Demonstrated on Real-world Electric Grid for Hurricane Impacted Outages

Patrick Maloney*, Xiaoyuan Fan, Marcelo Elizondo, Xue Li
Energy and Environment Directorate, Pacific Northwest National Laboratory, Richland, WA USA
*Corresponding author: patrick.maloney@pnnl.gov

*Abstract*—This work proposes a methodology for estimating recovery times for transmission lines and substations, and is demonstrated on a real-world 1269-bus power system model of Puerto Rico under 20 hurricane scenarios, or stochastic realizations of asset failure under the meteorological conditions of Hurricane Maria. The method defines base recovery times for system components and identifies factors that impact these base values by means of multipliers. While the method is tested on transmission lines and substation failures due to hurricanes, it is based on a generic process that could be applied to any system component or event as a general recovery time estimation framework. The results show that given the two failure modes under study (transmission towers and substations), transmission towers appear to have a greater impact on recovery time estimates despite substations being given longer base outage times. Additionally, average recovery times for the simulated hurricanes across 20 scenarios is ~28,000 work crew days.

**Sets**
$A$: Set of all asset types
$I$: Set of all system components
$F$: Set of all failure modes

**Indices**
$a$: Asset type $a \in A$ (Transmission towers, substations)
$i$: component $i \in I$
$f$: failure mode $f \in F$

**Parameters**
$X_{a,i,f} \in \{0, 1\}$ : Binary parameter indicates if asset type $a$, component $i$, has failed in mode $f$.
$OT_{a,f}$: Base outage time for asset type $a$, failure mode $f$ (units are work crew hours)
$Loc_{a,i}$: Location/terrain scalar for asset type $a$, component $i$
$Tech_{a,i}$: Technology scalar for asset type $a$, component $i$
$WC_{a,f}$: Distinct work crews available for parallel work on asset type a with failure mode $f$

## I. Introduction & Literature Review

Identifying resilience metrics has taken on greater importance as engineers continually try to improve responses to unexpected disasters. Of particular interest are long-term duration-based metrics designed to estimate recovery time as a way to augment the decision making of the public, policy makers, and disaster recovery experts during the planning, preparation, and responding stages of disaster response.

Existing reliability metrics such as Customer Average Interruption Duration Index (CAIDI) and System Average Interruption Duration Index (SAIDI) [1] often include durations in their definitions but do not attempt to estimate or refine these durations themselves. Rather they use empirical data. Similarly, recovery durations are discussed conceptually in terms of specific components of recovery time, but actual recovery time estimates are not computed. In [2], a resilience triangle is defined that conceptualizes the recovery from a disaster in terms of quality of infrastructure percentage. [3] [4] extend the resilience triangle to a resilience trapezoid with 3 states (post-event degraded state, post-restoration state, and resilient state) as well as transition times between states. Additionally, [4] estimates resilience-based metrics based on hurricane winds but does not attempt to quantify recovery durations.

Several research efforts attempt to compute actual recovery times. Reference [5] builds on [4] through duration-based resilience metric computations by using a mean time to repair (MTTR) of 10 and 50 hours for transmission lines and towers respectively, due to hurricane winds in Great Britain. Uncertainty associated with the actual Time To Repair (TTR) is modeled by associating the MTTR with an exponential distribution. In [6], averages and standard deviations of distribution system component outages are evaluated with the purpose of providing data to be used in predicting future outage durations for different sized events. In [7], percentage of load remaining is plotted as a function of hours since an event for multiple scenarios concerning earthquakes on the Portland General Electric system.

The contributions of this work include:

1. a general recovery time estimate (RTE) metric applicable to any number or type of system components, failure modes, or disasters requiring recovery.

2. a demonstration of the metric implemented for substation damage due to flooding and transmission line outages due to tower damage from high wind speeds on a realistic 1269-bus Puerto Rican power system model across 20 simulated hurricane scenarios. Here we define "hurricane scenarios" as stochastic realizations of asset failure under the meteorological conditions of Hurricane Maria.

The RTE is obtained by estimating base outage times for the failure modes of interest, and then identifying factors most likely to impact these base outage times which are implemented through scalar multipliers applied to the base outage times. Factors investigated for adjusting base recovery times in this work include location and technology. While the RTEs developed in this work are specific to substation and tower damage under hurricane scenarios, the framework used to develop these estimates is generalizable to any event, system component, or failure mode.

Furthermore, in addition to the value the RTE framework brings to estimating recovery times it can also serve as an important building block in planning and validating the sequence of asset recovery. For example, following a system event, experts supported by system analysts can build an asset recovery plan involving a sequence of components to recover serially or several sequences of components to recover in parallel. This can be based on expert knowledge of the system and may leverage power grid models to validate the recovery plan. Likely this will prioritize recovering high impact infrastructure with short recovery times. However, without a method to quickly estimate recovery times via a framework like the RTE proposed in this work, system analysts will not have the information they need to compare tradeoffs when determining asset recovery prioritization which may result in inefficient recovery times (Figure 1).

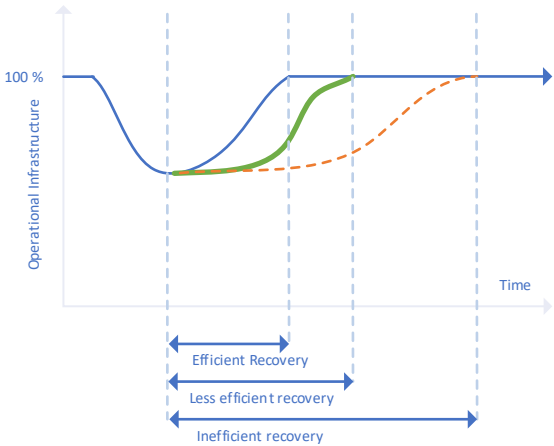

Figure 1: Resilience triangle adapted from [2]. Better estimates of component recovery times will allow analysts to plan and execute more efficient recoveries.

The remainder of this work is organized as follows: Section II describes the system under study and the simulation of hurricanes and power system failures. Section III introduces the RTE framework and how it is applied to the hurricane outage data obtained in Section II. Section IV discusses the application of the RTE framework to hurricane scenarios and Section V concludes and suggests future extensions.

## II. STUDY SYSTEM AND SIMULATION SETUP

The study system is a real-world 1,269-bus representation of Puerto Rico and the event under study is Hurricane Maria which occurred in 2017 and accounted for significant substation and transmission line failures. Using the fragility module in PNNL's Electrical Grid Resilience and Assessment System (EGRASS) [8], [9], twenty stochastic realizations of asset failure, or hurricane scenarios, are developed under the meteorological conditions of Hurricane Maria [10]. The hurricane scenario design method is based on the limitations discussed in [8]. Each hurricane scenario converts the fragility curve outage probabilities for substations and transmission towers into failure indicator statuses $X_{a,i,f}$ that can later be used in DCAT [8], [11] power systems studies. To convert each failure probability into an actual system state (failed or not failed), the failure probability is compared against a sampled value from a uniform distribution over the interval [0,1]. If the sampled value is less than the failure probability, the component is modeled as failed, otherwise the component remains operational.

The resulting impact of load being shed through time for each hurricane scenario based on DCAT results is given in Figure 2, where each of the twenty hurricane scenarios is represented by a unique color. Here Contingency Number refers to discrete snapshots in time of the power system as the hurricane progresses over it. Higher contingency number indicates time periods later in the hurricane's progression where more equipment has been outaged. Outaged equipment is observed indirectly in Figure 2 by total load not served, where larger unserved load indicates the impact of accumulated outages on the Puerto Rican power system. The final unserved load (rightmost point) for each hurricane scenario is the same, ~1,500 MW of system load not being served, resulting in a complete blackout. However, the exact equipment causing each blackout and the time (contingencies) it takes to reach this state are not identical across scenarios. To be clear, Figure 2 is the result of power system outages developed in EGRASS and used by DCAT power system simulations as in [8]. It is not the result of the analysis proposed in this paper. Rather this paper analyzes the outaged components which cause the blackouts in each scenario.

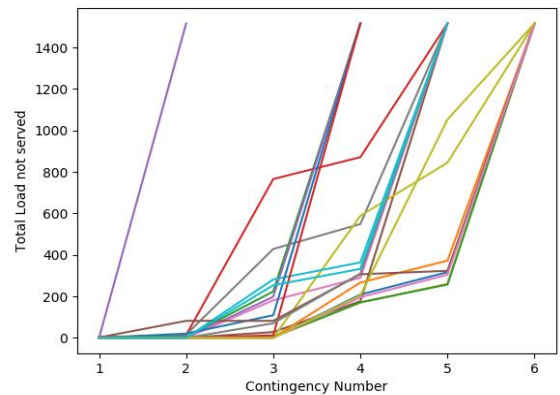

Figure 2: Total Load not served as function of contingency number in hurricane progression, each hurricane realization is represented by a unique color.

In terms of the electric power grid, a substation failure or outage means its corresponding system bus is out of service or de-energized. Any load connected to that bus is out of service and all generators connected to the bus are out of service. With regard to a transmission lines, a single tower failure takes the entire line that it supports out of service. Furthermore, the RTE of a specific transmission line is the sum of its tower RTEs. However, for the purposes of estimating RTEs of individual power systems components, and approximating the system recovery time, DCAT simulations are unnecessary. The only data needed comes directly from EGRASS which converts substation and tower failure probabilities into failure indicator statuses $X_{a,i,f}$ for each hurricane scenario.

While a statistical analysis of RTE could be accomplished directly from the probabilities established by the fragility curves contained in EGRASS, a disadvantage of analyzing RTE this way is that the statistics wouldn't correspond to any specific power system state. For example, an average RTE could be calculated from probabilities alone, but the RTE could not be related to an actual power system state with specific locations of the failed components. From a power systems analysis perspective, knowing this is useful as key grid health metrics such as energy not served can vary significantly depending on which components are out of service. However, one potential disadvantage of simulating hurricane scenarios is the required computational effort, as well as a large set of scenarios required for its statistics to converge and stabilize.

### III. RECOVERY TIME ESTIMATE (RTE) FRAMEWORK

The RTE for single hurricane scenario is defined by eq. (1), as formulated below:

$$RTE = \sum_{a,i,f} \frac{Loc_{a,i} * Tech_{a,i} * X_{a,i,f} * OT_{a,f}}{WC_{a,f}}. \quad (1)$$

The purpose of (1) is to provide an estimate of the time to recover all damaged system components for a given hurricane where the summation is over all system components. Factors in the numerator allow the recovery time for any element to be adjusted up, while factors in the denominator scale individual element recovery times down. The RTE as proposed would provide a utility, system operator, or government officials a first estimate of the time needed to recover a system following a disaster.

The RTE in (1) contains a single summation where $OT_{a,f}$ gives a base outage time dependent on the asset type and failure mode that might occur in units of work crew days (needed to restore the component). Work crew days denotes the number of days needed for a work crew properly equipped to repair the failure mode of the component. $OT_{a,f}$ is estimated for the metric in Table 1 where the transmission $OT_{a,f}$ is based on [12] and the station $OT_{a,f}$ is assumed to be 50% longer that the transmission $OT_{a,f}$. The RTE is a sum over all components belonging to each asset type and failure modes associated with each asset type in the system under study. Only those assets with a non-zero failure indicator parameter ($X_{a,i,f}$) contribute work crew days to the RTE.

Additionally, there are 2 scaling parameters that adjust the work crew days based on location or terrain ($Loc_{a,i}$) and technology ($Tech_{a,i}$) given in Table 2 and Table 3 respectively. The location scaling parameter is used to account for the additional time it takes to repair assets in unlevel terrain where it may be difficult to efficiently deploy construction equipment such as cranes and trucks. In the present studies this is only applied to transmission towers which commonly are sited on uneven terrain, whereas substations, which tend to be sited on level terrain use $Loc_{a,i} = 1$. Transmission tower $Loc_{a,i}$ values are determined by the terrain slopes which may be obtained from EGRASS. The technology scaling parameter $Tech_{a,i}$ is used to account for the additional time it takes to repair different types of assets. For example, a 230 kV tower foundation likely requires more concrete and thus more time to prepare than a 115 kV tower foundation, this should be reflected in the overall estimate if an entire transmission tower structure needs to be replaced. Similarly, a higher voltage substation will tend to have larger and more expensive components that a lower voltage substation, and the difficulty in working with these larger components is accounted for by $Tech_{a,i}$.

TABLE 1: $OT_{a,f}$ BASE OUTAGE PARAMETERS

| Description | Transmission Towers | Stations |
|---|---|---|
| $OT_{a,f}$ | 10 days | 15 days |

TABLE 2: TRANSMISSION TOWER $Loc_{a,i}$ MULTIPLIERS AND THRESHOLDS

| Description | $Loc_{a,i}$ | Thresholds |
|---|---|---|
| Trans. tower low slope | 1.0 | slope < 25° |
| Trans. tower moderate slope | 1.05 | 25° ≤ slope < 35° |
| Trans tower high slope | 1.1 | 35° ≤ slope |

TABLE 3: TRANSMISSION TOWER AND STATION TECHNOLOGY MULTIPLIERS AND THRESHOLDS

| Description | $Tech_{a,i}$ | Criteria |
|---|---|---|
| Base voltage equipment | 1 | nominal kV < 138 kv |
| Higher voltage equipment | 1.2 | nominal kV ≥ 138kv |

Finally, there is a scaling parameter $WC_{a,f}$ for the number of fully equipped work crews available for parallel work efforts on asset type $a$ under failure mode $f$. A fully equipped crew represents the correct number of people and equipment such as trucks, tools, cranes necessary to repair asset type $a$ under failure mode $f$. To prevent multiple work crews from working on the same asset type $a$, failure mode $f$, and component $i$ simultaneously, an assumption, (2), should be made that there will never be more crews outfitted for failure mode $f$ on asset type $a$ than there are of those types of failures existing concurrently in the field.

$$WC_{a,f} \leq \sum_{i,f} X_{a,i,f} \quad (2)$$

In other words, if two (components, i) transmission towers (asset type, a) are knocked over by hurricane winds (failure mode, f), then no more than two standard work crews $WC_{a,f}$ equipped to recover transmission towers (asset type, a) for hurricane winds (failure mode, f) should be dispatched to recover the 2 assets. Further work crews dispatched will result in idle time (e.g., only one person can operate a wrench at a time). For the results of this work $WC_{a,f}$ is set to 1. Thus the results of Section IV give total work crew days but can be approximately scaled by dividing by the number of work crews available to give more realistic recovery times.

As given, RTE described in (1)-(2) is extremely generic, readily expandable to additional base outage time multipliers like $Tech_{a,i}$ and $Loc_{a,i}$ in (1). Consequently, (1)-(2) may serve as a general RTE framework that could be applied to a variety of events and component failures rather than just hurricanes and substation and transmission outages. Presently, while the parameters $Tech_{a,i}$ (Table 2), $Loc_{a,i}$ (Table 3), and $WC_{a,f}$ ($WC_{a,f}$ equal to 1 in all results) use assumed values to develop and demonstrate the RTE framework, future work will seek to improve these parameters estimates. Factors that the scalar multipliers $Tech_{a,i}$, $Loc_{a,i}$, and $WC_{a,f}$ likely strongly depend upon are given in Table 4.

TABLE 4: FACTORS IMPACTING THE RTE SCALAR MULTIPLIERS

| Factors Impacting $Tech_{a,i}$ |
|---|
| (Transmission) |
| • kV class & tower design |
| • Number of circuits (e.g., single, double, etc.) |
| • Conductor type (ASCR, AAAC, etc.) |
| |
| (Substations) |
| • KV class and architecture |
| **Factors Impacting $Loc_{a,i}$** |
| • Access roads exist |
| • Urban vs. rural |
| • Flat vs. mountain |
| • Helicopter access only? |
| • Distance to service depots |
| **Factors Impacting $WC_{a,f}$** |
| • Available maintenance crew |
| • Crane availability |
| • Water pump availability |
| • Maintenance Budget |

IV. RESULTS & DISCUSSION

In this section we show the result of applying the RTE to 20 hurricane scenarios. Figure 3-Figure 4 show substation and transmission RTEs respectively. These figures show that while both substations and transmission towers impact the RTE, transmission outages are contributing significantly more to the RTE than substations despite substations being given longer base recovery times than towers (Table 1).

Additionally, while there appears to be some correlation between the magnitude of substation and transmission RTE values, there are exceptions. For example, the shape of Figure 3 and Figure 4 are similar for scenarios 8-20. However, scenarios 1,3,5,6 show that having a higher (or lower) RTE associated with substations does not always result in a higher (or lower) RTE associated with transmission lines.

In Figure 5, a box and whisker plot is given to demonstrate the high level of variation in transmission line RTE's based on cumulative transmission line tower failures for each line. In these plots the boxes cover quartiles 2 and 3 of the computed transmission line outages while the whiskers extend to no more than 150% of Q3-Q1 [13]. Because transmission line outages are based on a statistical realizations of their of their individual tower outages, the range of recovery time for them is much larger than for the substation failures modeled in this work. As observed, at least 1 line is causing approximately 2,000 work crew hours of maintenance in most scenarios. Given the base outage time for transmission towers is 10 days (Table 1), this specific line appears to frequently be the victim of ~200 tower failures. However, this line is outlier of the data and most boxes indicate that a majority of line RTEs are under 1,000 work crew hours of maintenance.

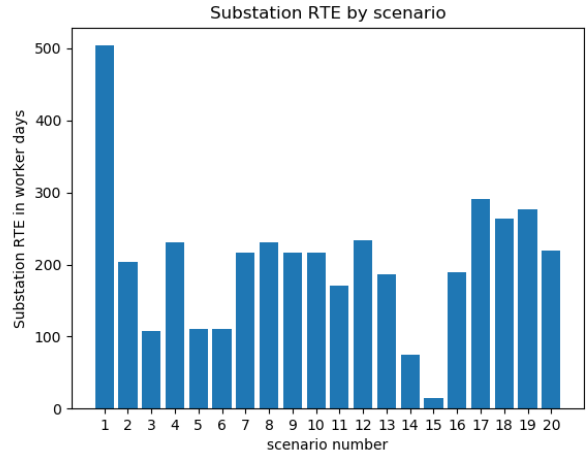

Figure 3: Total substation RTE by scenario

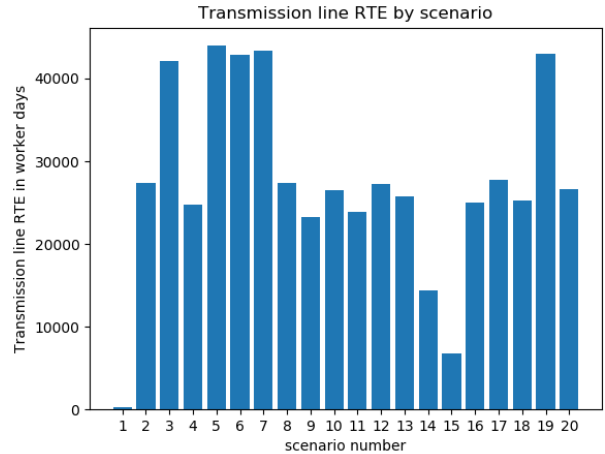

Figure 4: Total transmission RTE by scenario

Finally, in Table 5 we take the average and standard deviation of work crew days across all scenarios to illustrate the range and variation in the RTE across all 20 hurricane scenarios. In doing so we observe an overall average RTE of ~27,000 work crew days with standard deviation of ~12,000 work crew days. Future efforts with this work will seek to further increase the number of simulated hurricane scenarios so as to obtain steady statistics for scenario RTEs. This will also help determine the range the RTEs may occur over as a result of the simulated hurricanes.

In addition to the RTE framework computing outage times, future work may also use it to tune power grids to an appropriate level of resilience. For example, if a system is suspectable to a specific failure mode for a given event, such as transmission tower failures under hurricane scenarios, then it may be prudent to harden the transmission towers to this failure mode. This in turn will affect the fragility curves used to compute probability failures of the different components under study. Specifically, it should result in lower failure probabilities and ultimately lower cumulative RTEs when evaluated across many scenarios. By iterating between hardening components with high RTEs and simulating the impact of these decisions under hurricane scenarios, analysts will be able to tune grid design decisions to balance system cost and resilience.

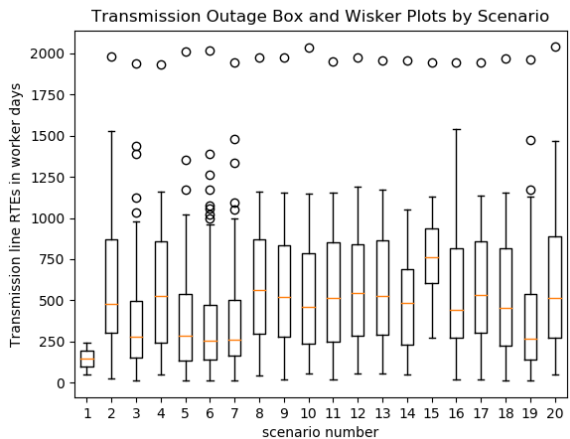

Figure 5: Transmission outage box and whisker plot.

TABLE 5: RECOVERY DAYS ACROSS ALL SCENARIOS.

| Component | Average (days) | Standard Deviation (days) |
|---|---|---|
| Transmission | 27,392 | 11,710 |
| Substation | 203 | 100 |
| Cumulative | 27,596 | 11,685 |

## V. CONCLUSIONS

Natural disasters can cause catastrophic damage to power system transmission and distribution networks. To properly estimate the energy not served from such extreme events, a method for estimating outage time due to transmission and substation failure is needed. In this work, we propose a methodology for estimating recovery times for transmission lines and substations. It is then demonstrated on a real-world 1269-bus power system model using stochastic realizations of asset failure under the meteorological conditions of Hurricane Maria. Simulation results show that transmission tower failures appear to have a greater impact on recovery time estimates compared to substation failures despite stations being given longer base recovery times than transmission towers. Additionally, average recovery times across all 20 hurricane scenarios is about 28,000 work crew days.

Potential future work includes expanding the RTE metric parameter space, validating the proposed RTE metric by comparing its results against real historical or simulated events simulating recovery times, and using the RTE to tune system resilience-based design decisions. Feedback from frontline utility engineers will also be collected and integrated, so it can be seamlessly adopted in grid asset management, operation, and planning applications.

ACKNOWLEDGMENT

This work is supported by the U.S. Department of Energy and Federal Emergency Management Agency. Pacific Northwest National Laboratory(PNNL) is operated by Battelle for the DOE under Contract DOE-AC05-76RL01830. The authors would like to thank LUMA Energy & PREPA for their involvement.